\documentclass[conference]{IEEEtran}
\IEEEoverridecommandlockouts
\usepackage{graphicx,cite,calc,xcolor,subfigmat,amssymb,amsmath,mathrsfs,dsfont,hyperref,epstopdf}
\usepackage[normalem]{ulem}

\usepackage{float}

\usepackage{soul} 
\usepackage{xcolor}
\usepackage{optidef}
\usepackage{listings}
\usepackage{enumitem}
\usepackage{hyperref}
\makeatletter
\newcommand*{\rom}[1]{\expandafter\@slowromancap\romannumeral #1@}
\makeatother
\usepackage { amsmath }
\usepackage{newtxtext,newtxmath}
\title{Power Allocation for a HAPS-Enabled MIMO- NOMA System with Spatially Correlated Channels}

\begin{document}

 \author{\rm{Rozita Shafie}, \rm{Mohammad Javad Omidi}, \rm{Omid Abbasi}, \rm{Halim Yanikomeroglu}
 \thanks{This work was supported in part by Huawei Canada Co., Ltd.}
 \thanks{Rozita Shafie and  Mohammad Javad Omidi are with the Department of Electrical and Computer Engineering, Isfahan University of Technology, Isfahan 84156-83111, Iran; Omid Abbasi and Halim Yanikomeroglu  are with the Department of Systems and Computer Engineering, Carleton University, Ottawa, ON K1S5B6, Canada.;
email: \texttt{shafie@ec.iut.ac.ir, omidi@iut.ac.ir , omidabbasi@sce.carleton.ca, halim@sce.carleton.ca }.}
}

\maketitle

\begin{abstract}
High-altitude platform station (HAPS) systems are considered to have great promise in the multi-tier architecture of the sixth generation (6G) and beyond wireless networks.
A HAPS system can be used as a super macro base station (SMBS) to communicate with users directly since  there is a significant line-of-sight (LoS) link between a HAPS and terrestrial users.
One of the problems that  HAPS SMBS systems face, however, is  the high spatial correlation  between the channel gain of adjacent users, which  is due to the LoS link between the HAPS and terrestrial users.
In this paper, in addition to utilizing the spatial correlation of channel gain between multiple users to improve user services, we consider correlated channel gain for each user. In the proposed method, terrestrial users with a high spatial correlation between their LoS channel gain  are grouped into  NOMA clusters. Next, an algorithm is proposed to allocate power among terrestrial users to maximize the  total rate  while satisfying the quality-of-service (QoS) and successive interference cancellation (SIC) conditions. Simulation results show that a HAPS SMBS has superior data rate and energy efficiency in comparison to a terrestrial BS.
\end{abstract}


\section{Introduction}
Ubiquitous super high data rate coverage is one of the key goals of the sixth-generation (6G) and beyond networks. High-altitude platform station (HAPS) systems have been proposed as a feasible technology to achieve ubiquitous super high data rate coverage in 6G network architecture \cite{9380673}, \cite{9772280}.
The advantage of a HAPS, in comparison to terrestrial and satellite communications layers, is that it can be used as a super macro base station (SMBS) to cover a large metropolitan area\cite{9380673}, \cite{9356529}.
Classical multiple-input multiple-output (MIMO) cannot be applied to a HAPS system directly because the line-of-sight (LoS) links between a HAPS and terrestrial users create a significant correlation between the channel gain of users\cite{7841647}.
Furthermore, in the literature on HAPS systems (e.g., \cite{DBLP:journals/corr/abs-2201-07379}), it is assumed that the emitted signals are uniformly distributed in the environment, and consequently, they consider uncorrelated fading for the channels of the users.
Nevertheless, there may not be a  rich scattering environment in a HAPS network, and the emitted signals may have more multipath components  from some spatial directions compared to others, and hence the practical channels are spatially correlated \cite{8861014}, \cite{3gpp.36.331}.


In a non-orthogonal multiple access (NOMA) scenario, users within the same cluster receive the same signal from the base station (BS) in downlink mode. We propose to solve the problem of spatial correlation between adjacent users by utilizing a NOMA scenario. 

The authors in \cite{8301007} allocated the power of a BS among users and clusters to maximize the total rate of the system.
In their proposed algorithm, the power of the BS was first allocated among all of the users in the cell to satisfy the quality-of-service (QoS) condition, and then the remaining power was allocated to the user of each cluster with the highest channel gain. This is because these users do not receive interference from other users, and they also have the best channel gain to maximize the total rate.
The user of each cluster with the best channel gain causes interference for other users in the cluster, and for this reason, the allocation of the remaining power to this user violates QoS and successive interference cancellation (SIC) constraints for other users in the cluster.

In this paper, terrestrial users are served with a MIMO-NOMA HAPS system. The channel gain for each terrestrial user is assumed as a correlated Rician fading channel. In order to resolve the spatial correlation issues, we begin by clustering users on the basis of spatial correlation. Next, we propose a power allocation algorithm that aims to maximize the total rate under users’ QoS and SIC conditions. Finally, we compare the total rate and energy efficiency of the terrestrial BS and HAPS SMBS.

\section{System Model}

\begin{figure}[htbp]
\centerline{\includegraphics[width = 9.cm ]{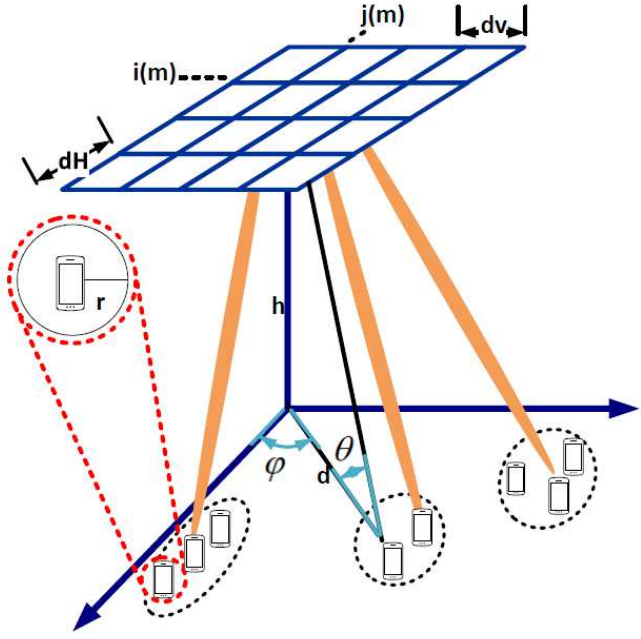}}
\caption{The proposed HAPS-enabled MIMO-NOMA system.}
\label{fig1}
\end{figure}

We begin by considering an integrated communication network that includes terrestrial users and a HAPS system, as depicted in Fig. \ref{fig1}. We utilize  a MIMO-NOMA HAPS system to transmit the superposed signal of terrestrial users in the downlink mode at the  GHz band. 
The HAPS is located at an altitude of approximately $20~\mathrm{km}$, and the transmitting user equipment (UE) is located in the terrestrial layer with a coverage radius of $1~\rm{km}$.
 The HAPS array consists of a uniform planar array (UPA) with $M_H$ horizontal antenna elements and $M_V$ antenna elements. The antenna spacing on horizontal and vertical elements are $d_H$ and $d_V$, respectively. 
 The total number of antenna elements is $M = M_V M_H$, and the antenna index of the UPA is  $m \in \left[ {1,M} \right] $. In general, The $m^{th}$ antenna is located as follows\cite{SIG-093}:
\begin{equation}
{\bf u}_m  = \left[ {\begin{array}{*{20}c}
   {i\left( m \right)d_H \lambda }  \\
   {j\left( m \right)d_V \lambda }  \\
   0  \\
\end{array}} \right],
\end{equation}
\begin{equation}
\begin{array}{l}
 i\left( m \right) = \bmod \left( {m - 1,M_H } \right), \\ 
 j\left( m \right) = \left\lfloor {{{\left( {m - 1} \right)} \mathord{\left/
 {\vphantom {{\left( {m - 1} \right)} {M_H }}} \right.
 \kern-\nulldelimiterspace} {M_H }}} \right\rfloor . \\ 
 \end{array}
\end{equation}
  The users are grouped into $M$ clusters, and they are  scheduled on a NOMA basis.  In this paper, the goal is to obtain the power allocation coefficient of the terrestrial users by maximizing the overall system rate. 
\subsection{Channel Model}
Since the  signal is not uniformly distributed  in a propagation environment, and also since there is an LoS path between the HAPS and terrestrial user, the channel for user $l$ in cluster $m$ is modeled with a Gaussian distribution as follows:
\begin{equation}
{\bf h}_{m,l}  \sim {\rm N} \left( {{\bf \bar h}_{m,l} ,{\bf R}_{m,l} }
\right),
\label{eq1}
\end{equation}
where $ {\bf \bar h}_{m,l} $ corresponds to the LoS component and  ${\bf R}_{m,l} $  is the positive semi-definite covariance matrix describing the spatial correlation of the NLoS components\cite{8620255}, \cite{8445853}.

In \cite{SIG-093} and \cite{7510943},  it was shown that only the LoS channel components cause the high spatial correlation between adjacent users, and that the correlation of NLoS components between users are sufficiently low  that they can  be ignored. Therefore, the spatial correlation is calculated only between the LoS components of the users' channel gains.
In the following, the LoS and NLoS channels are examined separately.

For an antenna array with $M$ elements, the LoS channel response ${\bf \bar h}_{m,l} \in {\bf C}^M  $ for  $l^{th}$ user in the $m^{th}$ cluster is given by:
\begin{equation}
{\bf \bar h}_{m,l}  = \sqrt {\beta _{m,l}^{\rm{LoS}} } \left[ {e^{j{\bf k}\left( {\varphi _{m,l} ,\theta _{m,l} } \right)^T {\bf u}_{\bf 1} } , \cdots ,e^{j{\bf k}\left( {\varphi _{m,l} ,\theta _{m,l} } \right)^T {\bf u}_{\bf M} } } \right]^T,
 \label{eqa2}
 \end{equation}
 where  ${\varphi _{m,l}} $ and $\theta _{m,l}$ are  the azimuth angle and the elevation angle of user $(m,l)$ to the HAPS antenna, respectively. 
 The wave vector  $ {\bf k}\left( {\varphi ,\theta } \right)$
 is also defined as follows\cite{SIG-093}:
\begin{equation}
{\bf k}\left( {\varphi ,\theta } \right) = \frac{{2\pi }}{\lambda }\left( {\begin{array}{*{20}c}
   {\cos \left( \theta  \right)\cos \left( \varphi  \right)}  \\
   {\cos \left( \theta  \right)\sin \left( \varphi  \right)}  \\
   {\sin \left( \theta  \right)} \\
\end{array}} \right).
 \end{equation}
 
 $ \beta _{m,l}^{\rm{LoS}} $ is the large-scale fading for the LoS path. The large-scale fading for the LoS and NLoS components in \rm{dB} are given by \cite{6863654}
\begin{equation}
\beta _{m,l}^{\rm LoS}  =  20\log d_{m,l} + 20\log f + 20\log \left( {\frac{{4\pi }}{c}} \right) + F_{m,l}^{\rm{LoS}} ,
 \end{equation}
\begin{equation}
\beta _{m,l}^{\rm NLoS}  =  20\log d_{m,l} + 20\log f + 20\log \left( {\frac{{4\pi }}{c}} \right)  + F_{m,l}^{\rm{NLoS}} ,
 \end{equation}
where $F_{m,l}  \sim N\left( {0,\sigma _{{\rm sf}}^2 } \right) $ is the shadow fading with standard deviations $ \sigma _{{\rm sf}}  = 1 $ for LoS and $ \sigma _{{\rm sf}}  = 20 $ for NLoS links, and ${d_{m,l} }$ is
the distance between user $l$ in the cluster $m$ and HAPS in meters. 

 Based on the 3D local scattering model  in \cite{SIG-093}, the $(a,b)^{th}$ element of  the spatial correlation matrix (${\bf R} \in C^{M \times M}$) is  given as follows\footnote{ The horizontal angular
spread $\left( {\Delta \varphi } \right)$ and the vertical spread $\left( {\Delta \theta } \right)$ 
are related to the 3D one-ring model in \cite{SIG-093}.
$\Delta \varphi  = \tan ^{ - 1} \left( {{r \mathord{\left/
 {\vphantom {r d}} \right.
 \kern-\nulldelimiterspace} d}} \right) $ and $\Delta \theta  = {{\left( {\theta _{\max }  - \theta _{\min } } \right)} \mathord{\left/
 {\vphantom {{\left( {\theta _{\max }  - \theta _{\min } } \right)} 2}} \right.
 \kern-\nulldelimiterspace} 2}$ that $\theta _{\min }  = \tan ^{ - 1} \left( {{h \mathord{\left/
 {\vphantom {h {\left( {d + r} \right)}}} \right.
 \kern-\nulldelimiterspace} {\left( {d + r} \right)}}} \right)$ and $\theta _{\max }  = \tan ^{ - 1} \left( {{h \mathord{\left/
 {\vphantom {h {\left( {d - r} \right)}}} \right.
 \kern-\nulldelimiterspace} {\left( {d - r} \right)}}} \right)$.  Also the value of elevation angle is given by $\theta  = {{\left( {\theta _{\max }  + \theta _{\min } } \right)} \mathord{\left/
 {\vphantom {{\left( {\theta _{\max }  + \theta _{\min } } \right)} 2}} \right.
 \kern-\nulldelimiterspace} 2}$. The parameter $r$ is the radius of the ring of local scatterers, and $ d$ is the horizontal distance of users.}: 
\begin{equation}
\left[ {\bf R} \right]_{a,b}  = \frac{{\beta ^{\rm{NLoS}} }}{{4\Delta \varphi \Delta \theta }}\int_{\theta  - \Delta \theta }^{\theta  + \Delta \theta } {\int_{\varphi  - \Delta \varphi }^{\varphi  + \Delta \varphi } {e^{j{\bf k}\left( {\varphi ,\theta } \right)^T \left( {{\bf u}_a  - {\bf u}_b } \right)} d\varphi d\theta .} } 
 \end{equation}

The probability of having an LoS path between an antenna element on the HAPS and a terrestrial user depends on the elevation angle and urban statistical parameters, and it can be formulated as follows \cite{6863654}:
\begin{equation}
P\left( {{\rm LoS}} \right) = \frac{1}{{1 + \kappa \exp \left( { - \omega \left( {\theta  - \kappa } \right)} \right)}},
 \end{equation}
Parameters 
$\kappa $ and $\omega $ are constants, depending on the environment. 
Subsequently, we also have $P\left( {{\rm NLoS}} \right) = 1 - P\left( {{\rm LoS}} \right)$ \cite{7881122}.

If the eigenvalue decomposition of ${\bf R} \in C^{M \times M} $ is given as $ {\bf R} = {\bf UDU}^H $, we can obtain the channel vector by using the Karhunen-Loeve expansion of ${\bf h}_{m,l} $, which is defined as follows\cite{8861014}:
\begin{equation}
{\bf h}_{m,l}  = {\bf \bar h}_{m,l}  + {\bf R}_{m,l}^{\frac{1}{2}} {\bf e} = {\bf \bar h}_{m,l}  + {\bf UD}^{\frac{1}{2}} {\bf U}^H {\bf e},
 \end{equation}
where ${\bf e} \sim N\left( {{\bf 0}_M ,{\bf I}_M } \right)$. 
 
  For simplicity, we write the above equations for a single antenna users,  but they are distributed to UEs with multiple antennas in the simulation results.
 
\subsection{User clustering in the downlink MIMO-NOMA system} Algorithm 1 presents a sub-optimal clustering method with low complexity that uses the correlation coefficients between the channel gain of multiple users.
Users with high spatial correlation are located in the same cluster.
In \cite{7841647}, it was proven that users with high spatial correlation have an adjacent location, and it is appropriate to use one emitted beam to serve them \cite{8653850}.
 The signal of users in each NOMA cluster is the same in the downlink mode, and for this reason, the high spatial correlation between the channel gain of users in each cluster does not decrease the favorable propagation. 
\begin{table}[htbp] 
\centering 
\begin{tabular}{l } 
 \hline\hline 
\textbf{Algorithm 1:} User-clustering for  MIMO-NOMA users.  \\ [0.5ex] 
\hline \hline 
\textbf{1.} Derive the LoS channel gain for each user.  \\ 
\textbf{2.} Calculate the correlation coefficient between different users as \\
\(R_{i,j}  = \frac{{\left\| {h_i h_j^t } \right\|}}{{\left\| {h_i } \right\|\left\| {h_j } \right\|}}\).\\ 
\textbf{3.} if  \(R_{i,j}  \ge \rho  \Rightarrow  \) 
put user i and user j  in the same cluster.\\[1ex]
  \hline 
  \end{tabular} \label{table:nonlin} 
 \end{table}

\section{Signal Processing}
The corresponding transmitted signals from the HAPS are modulated by a precoding matrix $ {\bf P} \in C^{M \times M}  $ and then transmitted over the radio channel.
The precoding matrix that is used by the BS in downlink mode is  indicated by $ {\bf P} \in C^{M \times M} $, which is equal to the identity matrix ($ {\bf P} = I_{_{M}}$) \cite{7236924,8301007}. 
Therefore, the transmitted  signals can be expressed as
\begin{equation}
 \begin{aligned}
{\bf x} = {\bf Ps},
\label{eq3}
\end{aligned}
\end{equation}
where the $ M \times 1 $ vector $ \bf s $ is given by \cite{8301007}
\begin{equation}
\begin{aligned}
{\bf s} =\left[ {\begin{array}{*{20}c}
   {\sqrt {P_{\max } \Omega _{1,1} } s_{1,1}  +  \cdots  + \sqrt {P_{\max } \Omega _{1,L} } s_{1,L} }  \\
    \vdots   \\
   {\sqrt {P_{\max } \Omega _{M,1} } s_{M,1}  +  \cdots  + \sqrt {P_{\max } \Omega _{M,L} } s_{M,L} }  \\
\end{array}} \right].
\label{eq4}
\end{aligned}
\end{equation}
Each element of the above matrix represents a superposed signal transmitted to the $ m^{th} $ cluster. 
 $\Omega _{m,l} $ denotes the NOMA power allocation coefficient for user $ (m,l) $, and $ s_{m,l} $ is the symbol of the  $ l^{th} $ user  in the $ m^{th} $ cluster.
The received signal for the $ l^{th} $ user of the $ m^{th} $ cluster can be expressed as follows:
 \begin{equation}
\begin{aligned}
{\bf y}_{m,l}  = {\bf H}_{m,l} {\bf Ps} + {\bf n}_{m,l},
\label{eq5}
\end{aligned}
\end{equation}
where $ {\bf n}_{m,l} $ is additive white Gaussian  noise with variance $ \sigma ^2$. 
The detection vector for  user $ (m,l) $  is represented by $ {\bf v}_{m,l}$. In order to completely remove the inter-cluster interference, the precoding and detection matrices need to satisfy  $ {\bf v}_{m,l}^H {\bf H}_{m,l} {\bf p}_k  = 0 $ for any $k \ne m $ \cite{8301007},
where $ {\bf p}_{k} $ is the $ k ^{th}$ column of $ {\bf P} $. 

After applying the detection vector $ {\bf v}_{m,l} $, the observed signal can be rewritten as follows \cite{8301007}:
\begin{equation}
\begin{array}{l}
 {\bf v}_{m,l}^H {\bf y}_{m,l}  = {\bf v}_{m,l}^H {\bf H}_{m,l} {\bf p}_m \sum\limits_{l = 1}^L {\sqrt {P_{\max } \Omega _{m,l} } } s_{m,l}  \\ 
  + \underbrace {\sum\limits_{k = 1,k \ne m}^M {{\bf v}_{m,l}^H {\bf H}_{m,l} {\bf p}_k {\bf s}_{\bf k} } }_{{\rm interference \: from \: other \: clusters}} + {\bf v}_{m,l}^H {\bf n}_{m,l}.  \\ 
 \end{array}
 \label{eq6}
\end{equation}
Since the inter-cluster interference is completely eliminated by using the detection matrix design, the equation \eqref{eq6} can be rewritten as follows:
\begin{equation}
\begin{aligned}
{\bf v}_{m,l}^H {\bf y}_{m,l}  = {\bf v}_{m,l}^H {\bf H}_{m,l} {\bf p}_m \sum\limits_{l = 1}^L {\sqrt {P_{\max } \Omega _{m,l} } s_{m,l} }   + {\bf v}_{m,l}^H {\bf n}_{m,l}.
\label{eq7}
\end{aligned}
\end{equation}
The value of $ \left| {{\bf v}_{m,l}^H {\bf H}_{m,l} {\bf p}_m } \right|^2  $ is defined as  the effective channel gain, and users in each cluster are arranged on the basis of it:
\begin{equation}
\left| {{\bf v}_{m,1}^H {\bf H}_{m,1} {\bf p}_m } \right|^2  \ge  \cdots  \ge \left| {{\bf v}_{m,L}^H {\bf H}_{m,L} {\bf p}_m } \right|^2.
\label{eq8}
\end{equation}
By applying the SIC to the received signal of each user, the interference is eliminated  from the users with lower channel gain. For this purpose, the following conditions must be satisfied\cite{7557079}:
\begin{equation}
\begin{array}{l}
P_{\max } \Omega _{m,l} \left| {{\bf v}_{m,l - 1}^H {\bf H}_{m,l - 1} {\bf p}_m } \right|^2  
- \qquad  \\
\quad  \sum\limits_{k = 1}^{l - 1}P_{\max } {\Omega _{m,k} } \left| {{\bf v}_{m,l - 1}^H {\bf H}_{m,l - 1} {\bf p}_m } \right|^2  \ge P_{\rm{tol}} , \; \; l = 2,3, \cdots ,L,
\label{eq9}
\end{array}
\end{equation}
where $ P_{\rm{tol}} $ is the minimum power difference required to differentiate the signal to be decoded and the remaining non-decoded signals\cite{7557079}. 
Therefore,  the user $ (m,l) $ sees interference from only users with greater channel gain than itself, and the data rate of user $ (m,l) $ is given by
\begin{equation}
\begin{aligned}
R_{m,l}  = \log _2 \left( {1 + \frac{{\rho \Omega _{m,l} \left| {{\bf v}_{m,l}^H {\bf H}_{m,l} {\bf p}_m } \right|^2}}{{1 + \rho \sum\nolimits_{k = 1}^{l - 1} {\Omega _{m,k}\left| {{\bf v}_{m,l}^H {\bf H}_{m,l} {\bf p}_m } \right|^2} }}} \right),
\label{eq10}
 \end{aligned}
 \end{equation}
where $ \rho  = \frac{{P_{\max } }}{{\sigma ^2 }} $.
\section{Problem Formulation}
The goal of this paper is to find the power allocation coefficients of MIMO-NOMA terrestrial users to maximize the total data rate. This problem can be formulated as follows: 
\begin{maxi!}|s|[2]
	{\Omega_{m,l} }{R^{\rm{sum}}\label{eq:ObjectiveExample3}}
	{\label{eq:Example3}}
	{}
\addConstraint{R_{m,l}}{\ge R_{_{m,l} }^{\rm{QoS}} ,  m \in \left\{ {1, \cdots ,M} \right\},  l \in \left\{ {1, \cdots ,L} \right\} \label{eq:constr-1}}
\addConstraint{P_{\max } \Omega _{m,l} \gamma _{_{m,l - 1}}  - P_{\max } \sum\limits_{k = 1}^{l - 1} {\Omega _{m,k} } \gamma _{_{m,l - 1}}}{\ge  P_{tol} \label{eq:constr-2}}
\addConstraint{P_{\max } \sum\limits_{m = 1}^M {\sum\limits_{l = 1}^L {\Omega _{m,l} } }}{ \le P_t,  \label{eq:constr-3}}
\end{maxi!}
where constraints \eqref{eq:constr-1}  and \eqref{eq:constr-2}  satisfy the QoS and SIC conditions of users, respectively, and \eqref{eq:constr-3} indicates the power limitation of the HAPS system.
The objective function of equation \eqref{eq:ObjectiveExample3} is non-convex, and hence obtaining an optimal solution for this problem is  non-trivial. In this paper, an iterative algorithm for solving the problem is proposed.
This algorithm consists of a primary part and a secondary part. To satisfy QoS and SIC constraints, the first algorithm allocates the minimum power to users. Then, the extra power of the HAPS is divided between users and clusters to maximize the sum rate of the system.
 
We obtain the minimum power allocation coefficients of MIMO-NOMA users to satisfy the QoS and SIC constraints in  \eqref{eq:constr-1} and \eqref{eq:constr-2} as follows:
\begin{equation}
 \Omega _{_{m,l} }^{\rm{QoS} }  = \left( {2{}^{R_{_{m,l} }^{\rm{QoS }} } - 1} \right)\left( {\sum\nolimits_{k = 1}^{l - 1} {\Omega _{_{m,k} }^{\rm{QoS} }  + \frac{1}{{\rho \left| {{\bf v}_{m,l}^H {\bf H}_{m,l} {\bf p}_m } \right|^2 }}} } \right) ,
\end{equation}
\begin{equation}
\Omega _{m,l}^{\rm{SIC}}  = \sum\limits_{k = 1}^{l - 1} {\Omega _{m,k}  + \frac{{P_{\rm{tol}} }}{{\rho \left| {{\bf v}_{m,l - 1}^H {\bf H}_{m,l - 1} {\bf p}_m } \right|^2 }}} .
\end{equation}
According to \eqref{eq:constr-3}, the power of the HAPS is limited, and in order to satisfy the QoS and SIC constraints of users,  the HAPS must have  a minimum amount of power, which is obtained from the following equation:
\begin{equation}
P_{\rm{req}}  = P_{\max } \sum\limits_{m = 1}^M {\sum\limits_{l = 1}^L {\Omega _{m,l}^{\min} } },
\end{equation}
where $ \Omega _{m,l}^{\min }  $  indicates the minimum power allocation of each user which is obtained in Algorithm 2.
\begin{table}[ht] 
\centering   
\begin{tabular}{  l }   
\hline \hline
\textbf{Algorithm 2:} Primary power allocation algorithm.
\\ 
 \hline \hline
\textbf{1. Initialization.} \\
 $  k = 1, R_{m,l}^{\min } ,P_{tol} ,\rho \left| {{\bf v}_{m,l}^H {\bf H}_{m,l} {\bf p_m} } \right|^2 ,l \in \left\{ {1, \cdots ,L} \right\}.$
\\ \\
\textbf{2. Calculate the QoS and SIC constraints.} \\
$  \Omega _{m,k}^{\rm{QoS}}  = \left( {2^{R_{m,k}^{\rm{QoS}} }  - 1} \right)\left( {\sum\limits_{l = 1}^{k-1} {\Omega _{m,l}^{\min } }  + \frac{1}{{\rho \left| {{\bf v}_{m,k}^H {\bf H}_{m,k} {\bf p}_m } \right|^2 }}} \right),  $\\
$ \Omega _{m,k}^{\rm{SIC}}  = \sum\limits_{l = 1}^{k-1} {\Omega _{m,l}^{\min } }  + \frac{{ P_{tol}  }}{{ \rho \left| {{\bf v}_{m,k-1}^H {\bf H}_{m,k-1} {\bf p}_m } \right|^2 }} \; \; for \; \; k\geq 2.$
\\ \\
\textbf{3. Satisfy both conditions simultaneously.} \\
$ \Omega _{m,k}^{\min }  = \max \left\{ {\Omega _{m,k}^{\rm{QoS}} ,\Omega _{m,k}^{\rm{SIC}} } \right\}.  $ \\[1ex]
\hline
\end{tabular} 
\end{table} 

In order to satisfy the SIC  and QoS constraints, we allocate power to users with Algorithm 2. In this way, the additional power of the HAPS can be allocated among users and clusters in order to maximize the overall system rate.
To this end, we first determine the power allocation within each cluster, and then we distribute power among clusters.

For power allocation  in each cluster, the user with the best channel gain maximizes the overall   rate of the system better than others. This is because this user does not experience interference from other users in the cluster, and it has a higher channel gain. For this reason, in the next algorithm, extra power is allocated to  the user with the best channel gain. Note that this user causes interference for other users in each cluster, and so when the extra power is allocated to the user with the highest channel gain, the QoS and SIC constraints of other users are violated. Therefore, at the same that we allocate extra power to users with the strongest channel gain in each cluster, the QoS and SIC of other users in that cluster must be updated.  

For power allocation among clusters, we use the following equation \cite{8301007}:
\begin{equation}
 \Delta P_{m,1}  = \left( {2^{\Delta R}  - 1} \right)\frac{{P_{\max } 2^{\sum\nolimits_{l = 1}^L {\hat R_{m,l} } } }}{{\rho \left| {{\bf v}_{m,1}^H {\bf H}_{m,1} {\bf p}_m } \right|^2 }}.
 \end{equation}
As we can see, if  $ \Delta P_{m,1} $ remains unchanged,  when $ \frac{{P_{\max } 2^{\sum\nolimits_{l = 1}^L {\hat R_{m,l} } } }}{{\rho \left| {{\bf v}_{m,1}^H {\bf H}_{m,1} {\bf p}_m } \right|^2 }} $
 is smaller, the data rate is higher.
 This idea can be used as a basis for the power allocation algorithm between clusters.
To this end, we first obtain each cluster's fraction amount ($ \frac{{P_{\max } 2^{\sum\nolimits_{l = 1}^L {\hat R_{m,l} } } }}{{\rho \left| {{\bf v}_{m,1}^H {\bf H}_{m,1} {\bf p}_m } \right|^2 }} $)  and arrange it in ascending order, starting with the lowest fraction level. Power is added to the fraction level until it reaches the next fraction level. Then, power is added to the first two fractions until they reach the third fraction. This process continues  until the extra power of the HAPS is finished.
\begin{table}[thb] 
\centering     
\begin{tabular}{l}   
 \hline
 \hline
\textbf{Algorithm 3:} Proposed power allocation algorithm.
\\ 
\hline \hline
\textbf{1.  Initialize parameters as follows:} \\
$ i=1, P_{\max } ,R_{m,l}^{\min } ,\rho \left| {{\bf v}_{m,l}^H {\bf H}_{m,l} \bf{p}_m } \right|^2 ,l \in \left\{ {1, \cdots ,L} \right\}. $\\
\\ 
\textbf{2. Sort clusters based on their fraction level.} 
\\
$ \left[ {H,m} \right] = {\rm{sort}}\left( {\frac{{P_{\max } 2^{\sum\nolimits_{l = 1}^L {R_{m,l}^{\min } } } }}{{\rho \left| {{\bf v}_{m,1}^H {\bf H}_{m,1} {\bf p}_m } \right|^2 }}} \right)$.\\
\\ 
\textbf{3.}  
${\bf while} \; ({\rm P}_{{\rm rem}}  > 0)$\\
$\left\{ {\Omega _{m\left( i \right),1}^{\min }  +  = \frac{{H\left( {i + 1} \right) - H\left( i \right)}}{{P_{\max } }}} \right.$\\
${\rm update}\left( {\Omega _{m\left( i \right),j}^{\rm{QoS}} ,\Omega _{m\left( i \right),j}^{\rm{SIC}} ,\Omega _{m\left( i \right),j}^{\min } } \right)$\\
$\left. {i +  = 1} \right\}$.\\ \\
\textbf{4. Calculate:}  \\
$R_{m\left( i \right),j}^{\min }  = \log _2 \left( {1 + \frac{{\rho \Omega _{m\left( i \right),j}^{\min } \left| {{\bf v}_{m\left( i \right),j}^H {\bf H}_{m\left( i \right),j} {\bf p}_{m\left( i \right)} } \right|^2 }}{{1 + \rho \sum\nolimits_{k = 1}^{j - 1} {\Omega _{m\left( i \right),k}^{\min } \left| {{\bf v}_{m\left( i \right),j}^H {\bf H}_{m\left( i \right),j} {\bf p}_{m\left( i \right)} } \right|^2 } }}} \right) \;$. \\[1ex]
\hline
\end{tabular} 
\end{table}

It should be noted that the fraction $ \frac{{P_{\max } 2^{\sum\nolimits_{l = 1}^L {\hat R_{m,l} } } }}{{\rho \left| {\rm v_{m,1}^H \rm H_{m,1} \rm p_m } \right|^2 }} $ is obtained by allocating power to the first user in the $m^{th}$ cluster. 
Adding power to only one user of the cluster violates the QoS and SIC conditions for the other users.  As discussed earlier, we allocate power to other users in the cluster as long as the QoS and SIC conditions of users are satisfied.
The summary of the allocation of the residual power to the maximum total rate with respect to QoS and SIC conditions of users is presented in Algorithm 3.

\section{Simulation Results}
In this section, numerical results are presented to  show the performance gain of the proposed scheme. 
The  simulation  parameters  are summarized in Table I. 
\begin{centering}
\begin{table}[h] 
\centering 
\begin{tabular}{c c } 
 \hline\hline 
\textbf{Table \rom{1}:} Simulation parameters. \\ 
  \hline \hline
   R (Area radius)  &  $ 1\; \rm{km} $  \\ 
     $f_c$  &  $ 2.5 \; \rm{GHz} $  \\  
     $ \rho $ & $ 0.7$\\
  $R_{\min}$ (minimum rate of each user ) &  $ 2\; {{\rm{bps}} \mathord{\left/
 {\vphantom {{\rm{bps}} {\rm{Hz}}}} \right.
 \kern-\nulldelimiterspace} {\rm{Hz}}} $  \\ 
   Thermal noise density & $  - 174 \; \rm{dBm/Hz}  $ \\
   $\left( {\kappa ,\omega } \right)$  & $ (9.61, 0.16) $ \\
   $ r$ (radius of the ring model around UE)  & $ 50 \;\rm{m} $ \\
  $ P_{\rm{tol}} $ & $ 1\; \rm{dBm} $ \\ [1ex] 
    \hline 
    \end{tabular} \label{table:nonl} 
    \end{table}
\end{centering}

\begin{figure}[htbp]
\centerline{\includegraphics[width = 9cm ]{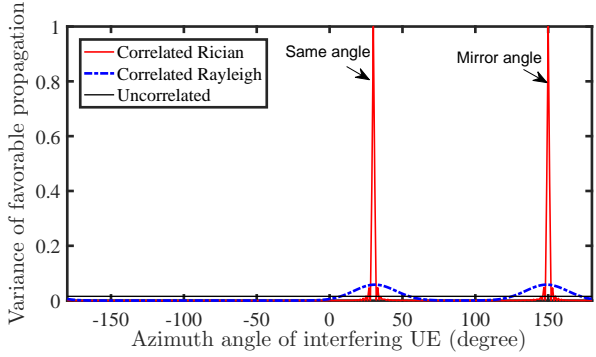}}
\caption{Variance of favorable propagation for $M=64$.}
\label{figurecorr}
\end{figure}

At first, we demonstrate the impact of the spatial correlation and LoS path on the favorable propagation. Fig. \ref{figurecorr} shows the variance of favorable propagation, i.e.,     
${\bf Var}\left\{ {{{{\bf h}_1 ^H {\bf h}_2 } \mathord{\left/
 {\vphantom {{{\bf h}_1 ^H {\bf h}_2 } {\sqrt {{\bf E}\left\{ {\left\| {{\bf h}_1 } \right\|^2 } \right\}{\bf E}\left\{ {\left\| {{\bf h}_2 } \right\|^2 } \right\}} }}} \right.
 \kern-\nulldelimiterspace} {\sqrt {{\bf E}\left\{ {\left\| {{\bf h}_1 } \right\|^2 } \right\}{\bf E}\left\{ {\left\| {{\bf h}_2 } \right\|^2 } \right\}} }}} \right\}$,
for correlated Rayleigh fading and correlated Rician fading channels. In Fig. \ref{figurecorr}, the azimuth angle of user 1 is $ {\raise0.7ex\hbox{$\pi $} \!\mathord{\left/
 {\vphantom {\pi  6}}\right.\kern-\nulldelimiterspace}
\!\lower0.7ex\hbox{$6$}}$, while the azimuth angle of user 2 varies between $ \left[ { - 180,180} \right]$ degrees.
For correlated channel gain, the variance of the favorable propagation varies with the angle of the UE, while it does not change for uncorrelated channel gain.
One can note that when  the azimuth angles of users are similar, there is a high spatial correlation between the channel gain of users,  and users cannot separate their signals easily. Based on this figure, it is clear that there is a high spatial correlation when there is an LoS path (i.e., a correlated Rician case) and hence the HAPS system is affected by this issue.  In order to mitigate this problem, users are grouped on the basis of spatial correlation, and we put the  correlated users  in the same cluster since the same signal is transmitted for them.

\begin{figure}[htbp]
\centerline{\includegraphics[width = 8cm ]{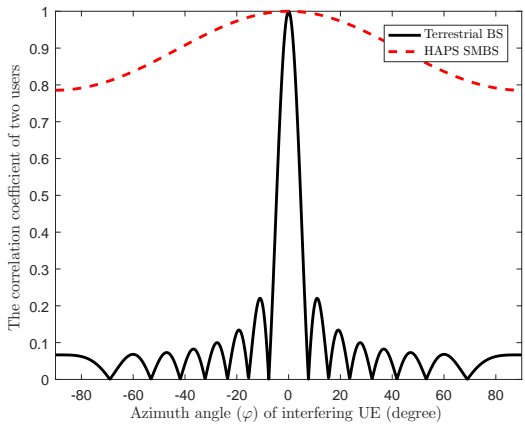}}
\caption{Correlation coefficient of two users for $M=100$.}
\label{locationcorr}
\end{figure}
  Fig. \ref{locationcorr} shows the importance of the elevation angle  for the spatial correlation in the HAPS system.
The correlation coefficient between the LoS channels of two users versus the azimuth angle of the interfering user is shown in this figure.
The azimuth angle and elevation angle of the first user are assumed to be zero degrees, and the azimuth angle of the second user changes between $\left[ {\frac{{ - \pi }}{2},\frac{\pi }{2}} \right]$. The elevation angle of the second user is calculated on the basis of the elevation of the BS in terrestrial and HAPS scenarios. The height of terrestrial BS is 25 meters whereas the height of HAPS is assumed to be $ 20\; \rm{km}$.
 Since the height of the HAPS antenna is larger than the terrestrial BS, the elevation angle  of the HAPS system is greater than the terrestrial BS. Therefore, the HAPS BS has a higher correlation coefficient in comparison to the terrestrial BS.

\begin{figure}[H]
\centerline{\includegraphics[width = 9cm ]{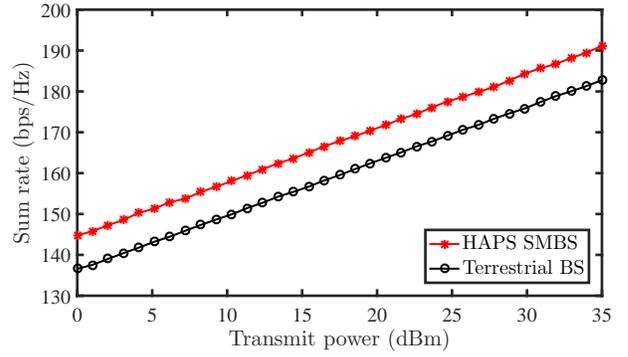}}
\caption{Sum rate of system versus the transmit power of BS for M=4.}
\label{SR_power}
\end{figure}
In Fig. \ref{SR_power}, we compare the performance of the proposed power allocation algorithm for the case where terrestrial users are served by the HAPS system and for the case where they are served by the terrestrial BS.
A vertical UPA antenna is assumed in the terrestrial BS whereas the UPA antenna on the HAPS is horizontal. 
Due to the terrestrial network propagation environment, channels between the terrestrial BS and users are assumed to be correlated Rayleigh fading channels, whereas correlated Rician channels are assumed between the HAPS and users. 
Fig. \ref{SR_power} thus shows the sum rate relative to the maximum power of the BS. The figure also shows that the HAPS has a much higher data rate in comparison to the terrestrial BS.
Indeed, since there is a direct link between the HAPS and terrestrial users, there is a high signal-to-noise ratio (SNR) for the downlink mode, and so the data rate of the HAPS system is greater than for the terrestrial BS.
\begin{figure}[htbp]
\centerline{\includegraphics[width = 9cm ]{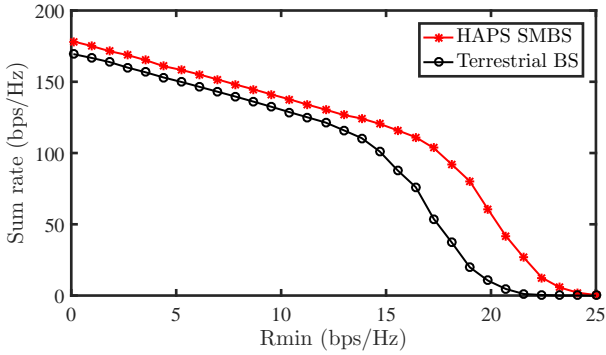}}
\caption{Sum rate versus the minimum rate per user for M=4.}
\label{SRRmin}
\end{figure}

In Fig. \ref{SRRmin}, the total rate of the HAPS and terrestrial BSs are depicted relative to the minimum rate of each user. As we can see, the total rate decreases with $R_{\min}$ because when the minimum rate for each user is large, to satisfy the QoS and SIC conditions, the BS has low residual power to maximize the total rate.

\begin{figure}[htbp]
\centerline{\includegraphics[width = 9cm ]{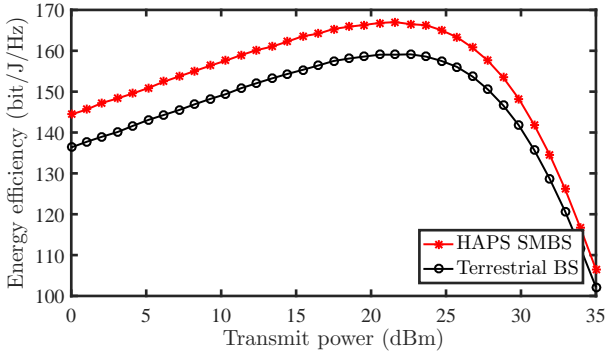}}
\caption{Energy efficiency versus the transmit power of BS for M=4.}
\label{power}
\end{figure}
Fig. \ref{power} compares the energy efficiency of the HAPS and terrestrial BSs  relative to transmit power. In line with previous figures,  the energy efficiency of the HAPS SMBS is greater than for the terrestrial BS.



\section{Conclusion}
In this paper, we considered the spatial correlation for a MIMO-NOMA HAPS system. Since the multipath distribution is not uniform in the propagation environment for a HAPS system, we modeled the channel gain as a correlated Rician fading channel. We also indicated that a high spatial correlation exists between the LoS paths of users. For this reason, terrestrial users with high spatial correlation were placed in the same cluster in the proposed method.
Next, we proposed an algorithm to allocate power among the users and clusters to maximize the total rate of the system so as to satisfy the QoS and SIC conditions. Simulation results showed that the HAPS SMBS has a superior data rate and energy efficiency compared to the terrestrial BS.
\ifCLASSOPTIONcaptionsoff
  \newpage
\fi
\bibliographystyle{IEEEtran}
\bibliography{IEEEabrv,Bibliography}
\vfill
\end{document}